\documentclass[oribibl, numbers]{llncs}

\usepackage[colorlinks = true,
  linkcolor = blue,
  citecolor = blue,
  urlcolor = blue]{hyperref}

\usepackage{url}
\usepackage{amsmath}
\usepackage{balance}
\usepackage{graphicx}
\usepackage{tabularx}
\usepackage{booktabs}
\usepackage{threeparttable}

\usepackage{fancyhdr} % for arxiv
\fancypagestyle{firststyle}
{
\fancyhf{}
\fancyfoot[C]{\scriptsize{Proceedings of the 37th International Conference on Testing Software and Systems (ICTSS 2025), Limassol, Springer, pp.~3--15. This version is the author's copy. The publisher's definite version is available online at Springer via \url{https://doi.org/10.1007/978-3-032-05188-2_1}.}}
}

\begin{document}

\title{A Time Series Analysis of Assertions \\ in the Linux Kernel}

\author{Jukka Ruohonen\orcidID{0000-0001-5147-3084}
\institute{University of Southern Denmark, \email{juk@mmmi.sdu.dk}}}

\maketitle

\begin{abstract}
Assertions are a classical and typical software development technique. These are
extensively used also in operating systems and their kernels, including the
Linux kernel. The paper fills a gap in existing knowledge by empirically
examining the longitudinal evolution of assertion use in the Linux
kernel. According to the results, the use of assertions that cause a kernel
panic has slightly but not substantially decreased from the kernel's third to
the sixth release series. At the same time, the use of softer assertion variants
has increased; these do not cause a panic by default but instead produce
warnings. With these time series results, the paper contributes to the existing
but limited empirical knowledge base about operating system kernels and their
long-term evolution.
\end{abstract}


\section{Introduction}

\thispagestyle{firststyle} % for arxiv

Assertions are a classical programming technique used to verify a program's
execution at runtime. Their history traces all the way back to the early works
by von Neumann and Turing~\cite{Clarke06}. In addition to verification,
assertions serve many other functions too. Among other things, they also serve
to document code for other developers, which can be particularly valuable when
operating with legacy systems and code~\cite{Hoare03}. Assertions are typically
used within functions to test preconditions placed at the beginning of a
function and postconditions placed at the end of the function, although
constraints can be placed also upon return values and intermediate states of the
function~\cite{Rosenblum95}. Assertions can be used also for testing many other
conditions, such as whether a condition or a state is satisfied or reached at
some time in the future~\cite{Bombieri25, Clarke06, Witharana25}. In software
engineering and the computing world in general, assertions are also behind a
classical debate between so-called defensive programming and design by contract.

In general, the defensive variant does not trust anything coming from outside,
thus seeking to particularly validate each and every input~\cite{Teto17}.  In
contrast, the design by contract programming philosophy assumes an implicit
contract made explicit through assertions between a caller and a supplier; a
supplier can assume that certain conditions are always guaranteed by a caller,
whereas the caller can assume a safe output provided by the supplier. Some
implementations have made the implicit contract visible by using a keyword
\texttt{assume} for precondition checks and a keyword \texttt{promise} for
postcondition checks~\cite{Rosenblum92}. Given such checks, ``any runtime
violation of an assertion is not a special case but always the manifestation of
a software bug'', tracing of which can be started by examining the given
preconditions, implying ``a bug in the client (caller)'', and the given
postconditions associated with ``a bug in the supplier (routine)''
\cite[p.~44]{Meyer92}. Regardless whichever side one takes in this classical
debate, assertions are frequently used---and triggered---also in operating
systems and their kernels. The use is presumably explained partially by the
context: given the overhead caused by defensive programming practices
particularly for resource-constrained systems, assertions have the obvious
benefit that these can be turned off during compilation~\cite{Clarke06}. This
background motives the paper's focus on the Linux kernel.

With respect to the Linux kernel, assertions have frequently been observed to be
behind many of crashes triggered by kernel fuzzing
frameworks~\cite{Ruohonen19RSDA}. In the kernel context an explicitly triggered
crash corresponds with a kernel panic. Without an assertion causing a panic, an
alternative behavior might be that a whole operating system just hangs. Also
these cases have historically been common in the Linux kernel~\cite{Gu03,
  Yoshimura12}. Thus, the design by contract philosophy has a slightly different
interpretation in the kernel context within which failing spectacularly may
often, or at least sometimes, be a better option than to continue execution in
undefined waters~\cite{FreeBSD24, Hoare03}. Different fallback and recovery
solutions are often also difficult to design and implement in the kernel
context.

These and related points motivated a lively debate recently among kernel
developers and other open source developers. The motivating argument put forward
was that Linux kernel developers should slowly move away from using assertions
that cause an outright panic; instead, they should prefer warnings and try to
come up with means to recover from fatal errors~\cite{LWN24a}. In the Linux
kernel the former panic-inducing assertions are used with different
\texttt{BUG()}-style macros, whereas the latter are typically used with
different \texttt{WARN()}-style variants. For a lack of a better terminology,
these are hereafter known as ``hard'' and ``soft'' assertions. Regarding the
hard assertions, the kernel documentation argues that ```I'm too lazy to do
error handling' is not an excuse for using \texttt{BUG()}'', and that major
``internal corruptions with no way of continuing may still use \texttt{BUG()},
but need good justification''~\cite{Kernel24a}. With this guidance and the
debate in mind, the following two research questions~(RQs) are~examined:
\begin{itemize}
\itemsep 3pt
\item{RQ.1: Has the kernel's growing size been met with a growing use of soft
  assertions in some or all of its notable subsystems?}
\item{RQ.2: Has the amount of hard assertion declined over time in some or all of the notable subsystems of the Linux kernel?}
\end{itemize}

Regarding the paper's structure, there are four sections remaining. The first of
these, Section~\ref{sec: related work}, briefly discusses related work from a
broader perspective. Then, Section~\ref{sec: setup} elaborates the research
setup, including the empirical data used and the methods for its
analysis. Results are presented in Section~\ref{sec: results}. The final
Section~\ref{sec: conclusion} concludes.

\section{Related Work}\label{sec: related work}

In addition to the works already referenced, a couple of larger knowledge bases
are worth mentioning. These allow to also better motivate the RQs. Thus,
regarding RQ.1, the software evolution research domain provides a good reference
point. According to the so-called laws of software evolution, which were
developed by Lehman and associates from the 1970s onward~\cite{Lehman97},
\text{many---if} not the most---software artifacts are continuously changing,
continuously growing, and continuously getting more
complex~\cite{Ruohonen15JSEP}. These laws correlate with the more recent concept
of technical debt; it is necessary to pay back maintenance debt in order to
continue to have a given software artifact in a reasonable state in the face of
continuous growth and increasing complexity. The Linux kernel has been seen as a
prime example supporting the Lehman's laws~\cite{Israeli10}. Regarding growth,
particularly new device drivers continuously added have been the driving
force~\cite{Chou01, Ruohonen24EASE}. Partially due to the growth, the Linux
kernel is also continuously and heavily refactored, as manifested by the
extensive amount of code churn~\cite{Ruohonen19RSDA}. In other words, technical
debt is also continuously paid back. These points provide a motivation also for
RQ.1. Although developers are encouraged to move away from hard assertions,
retaining the existing but perhaps implicit design by contract principle
requires that also soft assertions should grow alongside the growth of the
kernel's code~base.

The second knowledge base is the somewhat limited amount of existing empirical
research on assertions. This claim should not be interpreted to mean that
empirical research would be absent altogether; among other things, the benefits
of assertions in software testing have been investigated
empirically~\cite{Chen24, Taromirad24}. In this regard, assertions have been
observed to correlate with the effectiveness of test suites~\cite{Zhang15} and
less buggy files~\cite{Kudrjavets06}. Another empirical research area has
focused on the benefits of assertions for troubleshooting and debugging,
reusability, understandability, and documentation~\cite{Muller02, Patel17,
  Peruma24, Winkler24}. This research area aligns closely with the design by
contract philosophy. However, to the best of the author's knowledge, no
empirical studies have been conducted about the longitudinal evolution of the
use of assertion in the Linux kernel---or any other operating system kernel for
that matter. Thus, the paper fills a small but notable gap in existing software
engineering knowledge.

\section{Research Setup}\label{sec: setup}

\subsection{Data and Time Series}

The Linux kernel more or less follows the traditional semantic versioning
scheme. Release series are the highest level of abstraction; currently, the
Linux kernel operates with the sixth release series. Major releases, such as
$6.1, 6.2, 6.3, \ldots$, are made based on a loosely time-based release process;
a new major release occurs roughly every two or three
months~\cite{Kernel24b}. The adjective major signifies that such releases may
contain everything from new features to internal changes in the kernel's
application programming interface. Each major release typically changes several
hundred thousand lines of code (LOC). In addition, each major release is
maintained by a separate team who manages the release engineering of minor
releases, such as $6.1.1, 6.1.2, 6.1.3$, and so forth. These minor releases
typically contain only bug fixes. Therefore, the major releases are more
suitable for observing the evolution of assertions in the kernel.

Thus, each major release in the third, fourth, fifth, and sixth release series
was downloaded from the kernel's online archives~\cite{Kernel24c}. At the time
of writing (20 December 2024), the amounts of major releases released in these
release series were 20, 21, 20, and 13, respectively. Thus, the length of all
time series constructed is 74 releases. Although the length is rather short in
general, it is still sufficient for time series analysis with formal methodologies.

Both the overall software architecture of the Linux kernel and its software
development rely on different subsystems~\cite{Ruohonen24EASE}. In what follows,
the time series analysis is also based on major subsystems of the kernel. The
major ones observed are: \texttt{arch} (architectural code dependent on
instruction set architectures), \texttt{drivers} (device drivers), \texttt{fs}
(file systems), \texttt{net} (networking), and \text{sound} (sound
architectures). By following existing research~\cite{JiangJiang24,
  Ruohonen24EASE}, these were proxied by the top-level directories in the
kernel's file system representation. In addition, a category of \texttt{others}
was used to merge other notable subsystems, such as \texttt{kernel} for general
kernel code, \texttt{crypto} for cryptographic implementations, and
\texttt{security} for the kernel's security implementations. In addition, this
leftover category contains all other top-level directories insofar as they
contained C programming code, as identified by files ending to a
\texttt{.c}~suffix.

The reason for the leftover category (or, hereafter, a ``subsystem'' in order to
simplify the language) is that the kernel's subsystems vary considerably in
terms of their sizes. This point is evident from Table~\ref{tab: subsystems},
which shows the average sizes of the subsystems in terms of \texttt{.c}-files
contained in these and the total amount of LOC, including comments, in these
files, scaled by the seventy-four releases sampled. Even with the merging of
top-level directories, \texttt{others} is smaller than the other subsystems in
terms of LOC. The device driver subsystem is the largest one regardless of a
size metric.

\begin{table}[th!b]
\centering
\caption{Average Sizes of the Subsystems Across Releases}
\label{tab: subsystems}
\begin{tabular}{lcrcr}
\toprule
Subsystem\qquad\qquad\qquad\qquad && Files &\qquad\qquad& LOC \\
\hline
\texttt{arch} && $4,834$ && $1,590,037$ \\
\texttt{drivers} && $14,432$ && $11,692,942$ \\
\texttt{fs} && $1,224$ && $1,146,606$ \\
\texttt{net} && $1,321$ && $964,467$ \\
\texttt{sound} && $1,253$ && $931,181$ \\
\texttt{others} && $2,438$ && $82,114$ \\
\bottomrule
\end{tabular}
\end{table}

\pagebreak % arxiv
Then, for each subsystems, four time series were constructed:

\begin{enumerate}
\itemsep 5pt
\item{$\textmd{FILE}_{it}$ counts the number of files containing C programming
  code, as identified by a file's ending to a suffix \texttt{.c}, in the $i$:th
  subsystem of the $t$:th major release.}
\item{$\textmd{LOC}_{it}$ measures the lines of code, including code comments,
  of the files containing C code, as identified similarly to
  $\textmd{FILES}_{it}$, in the $i$:th subsystem of the $t$:th major release.}
\item{$\textmd{BUG}_{it}$ measures the amount of hard assertions in the C files
  present in the $i$:th subsystem of the $t$:th major release. These were
  identified by parsing each file line by line and counting the occurrences of the character strings \texttt{BUG(} and \texttt{BUG\_ON(}.}
\item{$\textmd{WARN}_{it}$ measures the amount of soft assertions in the
  $\texttt{.c}$-suffixed files present in the $i$:th subsystem of the $t$:th
  major release. Analogously to $\textmd{BUG}_{it}$, these were identified by
  parsing each file line by line and counting the occurrences of the character
  strings \texttt{WARN(}, \texttt{WARN\_ON(}, and \texttt{WARN\_ON\_ONCE(}.}
\end{enumerate}

To give a short example about the use of the assertion variants in practice, the
file \texttt{./drivers/base/regmap/regcache.c} triggers a kernel panic in a
function in case a precondition is not satisfied by a caller:

\begin{verbatim}
        void regcache_exit(struct regmap *map)
        {
                if (map->cache_type == REGCACHE_NONE)
                        return;

                BUG_ON(!map->cache_ops);
        [...]
\end{verbatim}

The above case is a good example because it could be argued that a softer
assertion variant might work or that an explicit check could be done analogously
to checking for the \texttt{REGCACHE\_NONE} constant. The same file uses also
\texttt{BUG()} as a \texttt{default} fallback case in a \texttt{switch}
statement, indicating for a caller that only the values explicitly listed in the
statement are assumed to be valid.

Then, the computations are carried out with the plain $\textmd{BUG}_{it}$ and
$\textmd{WARN}_{it}$ time series as well scaled variants of the two time
series. Two simple scaling functions are used: the first scales the plain series
by the total amount of hard and soft assertions in the $i$:th subsystems of the
$t$:th release, and the second by a subsystem's size at a given major
release. In other words:

\begin{equation}
\overline{\textmd{BUG}}_{it} = \frac{\textmd{BUG}_{it}}{\textmd{BUG}_{it} + \textmd{WARN}_{it}} \quad\textmd{and}\quad
  \widetilde{\textmd{BUG}}_{it} = \frac{\textmd{BUG}_{it}}{\textmd{FILE}_{it}},
\end{equation}
and likewise for $\textmd{WARN}_{it}$. The scaling by $\textmd{FILE}_{it}$ is
arbitrary; none of the results reported changed when scaled by
$\textmd{LOC}_{it}$. Finally, it can be remarked that a replication package is
available online~\cite{Ruohonen24rep}. It contains the dataset and the code.

\subsection{Methods}\label{subsec: methods}

Statistical methods for cumulative sum (CUSUM) processes are suitable for
examining the research questions. The history behind CUSUM processes is rich,
starting from the 1950s and peaking in the 1970s during which first formal
statistical tests were derived for testing structural changes in time
series~\cite{Brown75}. By a structural change it is meant that a model estimated
for a time series does not remain stable over time. To this end, a term
parameter stability is sometimes used as an alternative~\cite{Chen12}. The
concept is much broader than the concept of (covariance) stationary, which only
requires that a mean and a variance of a time series remain constant. However,
in the present context these two concepts are closely related because the focus
is on the means of $\textmd{BUG}_{it}$ and $\textmd{WARN}_{it}$. With respect to
RQ.2, a~weak requirement for a positive answer is that the means of the former
should \textit{not} remain constant for all or at least most $i$. A~strong
requirement implies that the means of both series should \textit{not} remain
constant in all or most subsystems because developers are encouraged to move
from hard assertions to the softer variants.

The CUSUM processes are thus computed for both ``static mean models'' and
``dynamic autoregression models'', both estimated with standard ordinary least
squares (OLS). To clarify these modeling concepts, the former models merely
estimate $\textmd{BUG}_{it} = \alpha_t + \epsilon_t$ and $\textmd{WARN}_{it} =
\beta_t + \varepsilon_t$, where $\alpha_t$ and $\beta_t$ are intercepts, and the
residuals $\epsilon_t$ and $\varepsilon_t$ are assumed to have zero means and
variances $\sigma^2_{\epsilon_t}$ and $\sigma^2_{\varepsilon_t}$ for $t = 1,
\ldots, 74$. The underlying (null) hypotheses are that $\alpha_1 = \alpha_2 =
\cdots = \alpha_{74}$ and $\sigma^2_{\epsilon_1} = \sigma^2_{\epsilon_2} =
\cdots = \sigma^2_{\epsilon_{74}}$, and likewise for $\beta_t$ and
$\sigma^2_{\varepsilon_t}$. In particular, the conditional means should remain
constant, although the word conditional is a slight misnomer in the present
context because only constants are included.

The dynamic first-order autoregression, denoted by AR(1), models are estimated
for differenced series, $\Delta\textmd{BUG}_{it} = \textmd{BUG}_{it} -
\textmd{BUG}_{it-1}$ and $\Delta\textmd{WARN}_{it}$, given that many of the
series are not stationary. For each subsystem \text{$i = 1, \ldots, 6$},
a~simple autoregression model $\Delta\textmd{BUG}_{it} = \phi_t + \rho_t
\Delta\textmd{BUG}_{it-1} + u_t$ is estimated, and likewise for
$\Delta\textmd{WARN}_{it}$. The underlying hypothesis again is that $\phi_t$,
$\rho_t$, and the variance $\sigma^2_{u_t}$ of the residual term $u_t$ remain
constant throughout the seventy-four major releases of the Linux kernel. The
same applies to the corresponding parameters in the AR(1) model for
$\Delta\textmd{WARN}_{it}$.

The CUSUM processes are particularly useful as visual diagnostic tools. The
\textit{strucchange} package~\cite{Zeileis02} for R is used for the
visualizations and associated computations. It allows to visualize the CUSUM
processes computed with visual boundaries; 99\% confidence level boundaries are
used for the visualizations presented. With the earlier points in mind and using
the \texttt{eventually} assertion notation proposed in the
literature~\cite{Bombieri25}, the weak requirement for a positive answer to RQ.2
means that \texttt{eventually("the boundaries are crossed" $\equiv$ TRUE)} for
all or at least most $i$. The logic is thus similar to using CUSUM processes for
anomaly detection~\cite{Olufowobi19}. In addition, the \textit{strucchange}
package offers more formal so-called M-fluctuation tests~\cite{Zeileis06}. These
are used to compute analogous models than the static (OLS) mean models used for
the visualizations except that a Poisson regression is used for the plain
$\textmd{BUG}_{it}$ and $\textmd{WARN}_{it}$ time series. The four scaled series
$\overline{\textmd{BUG}}_{it}$, $\widetilde{\textmd{BUG}}_{it}$,
$\overline{\textmd{WARN}}_{it}$, and $\widetilde{\textmd{WARN}}_{it}$ are tested
with OLS because they do not represent count data any~more.

In addition, to examine RQ.1 in a little more detail, a simple non-linear growth
curve estimation is carried out. Although numerous growth curves have been
specified in the literature~\cite{Ruohonen15COSE}, the standard logistic growth
curve is sufficient for the present purposes. It is written in the
\textit{growthcurver} package~\cite{Sprouffske16} used for computation as:
\begin{equation}
y_t = \frac{\alpha}{1 + \frac{\alpha - y_0}{y_0} e^{-\beta t}} ,
\end{equation}
where $y_t$ denotes a time series estimated with non-linear least squares,
$\alpha$ is the asymptotic maximum, $y_0$ is the initial value for a growth, and
$\beta$ is the growth rate. The interest is not to examine the statistical
performance of the logistic growth curves estimated, but rather to examine the
growth rate parameters across the time series. With respect to RQ.1, an
estimated $\hat{\beta}$ for $\textmd{WARN}_{it}$ should be roughly comparable to
equivalent parameters estimated for $\textmd{FILE}_{it}$ and $\textmd{LOC}_{it}$
in all $i = [1, 6]$ subsystems. In this case the kernel's pace of growth would
be matched by use of soft assertions at a roughly same speed. It should be
mentioned that some of the series are not increasing but decreasing---as is also
presumed by RQ.2. As the logistic growth curve is ill-suited for such series,
the growth rate coefficients are interpreted together with their standard~errors.

\section{Results}\label{sec: results}

\begin{figure*}[th!b]
\centering
\includegraphics[width=\linewidth, height=6.5cm]{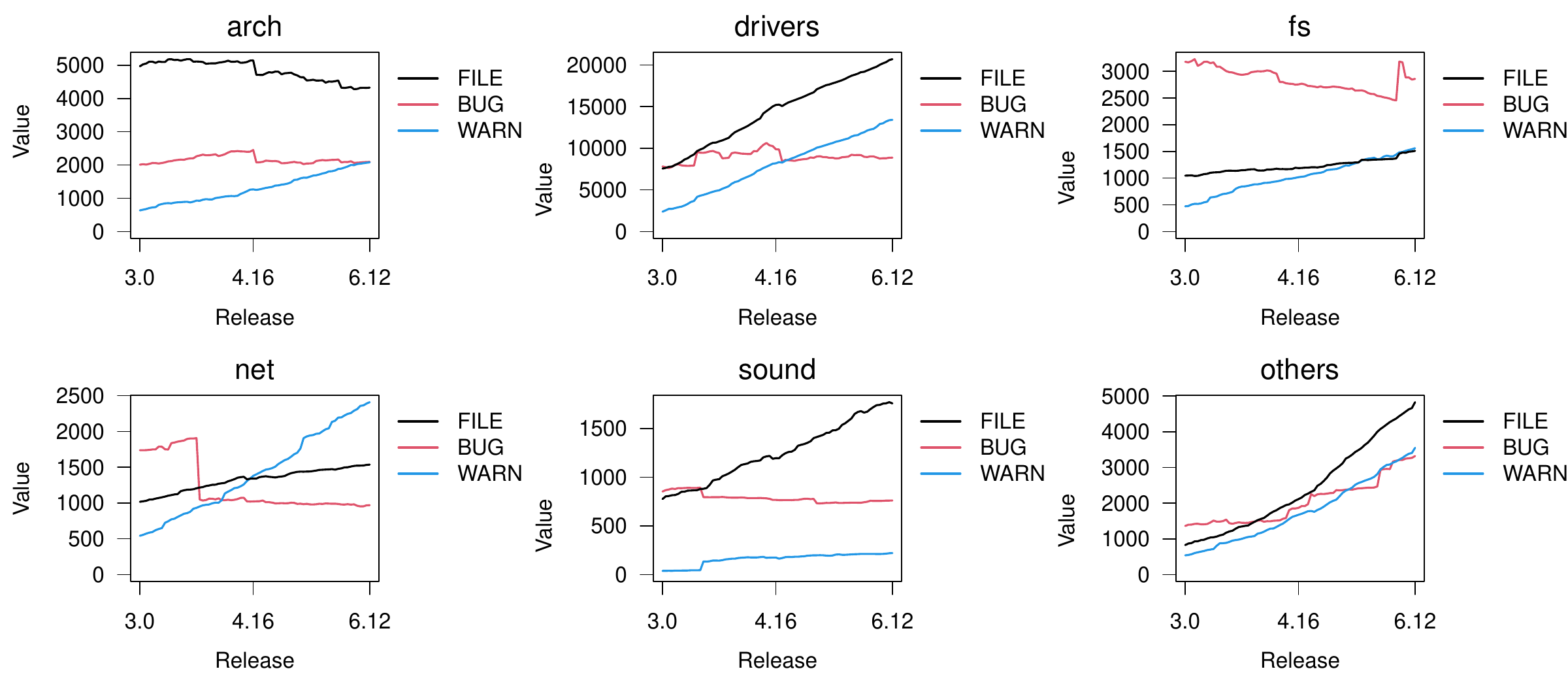}
\caption{$\textmd{FILE}_{it}$, $\textmd{BUG}_{it}$, and $\textmd{WARN}_{it}$}
\label{fig: series}
\end{figure*}

The presentation of the results can be started by taking a look at the time
series across the six subsystems. Three main series are thus shown in
Fig.~\ref{fig: series}. The conventional wisdom holds well: the \texttt{drivers}
subsystem has grown considerably in terms of new C programming files added
throughout the releases. Although the magnitudes are much smaller, also the
\texttt{sound} and \texttt{others} subsystems have witnessed accelerating
growth. Also the \texttt{fs} and \texttt{net} subsystems have grown, although at
a much slower pace and smaller magnitudes. The \texttt{arch} subsystem is the
only one in which the amount of \text{\texttt{.c}-suffixed} files has
decreased. Although a reason can be left for further examinations, the
observation could be interpreted in a positive light in a sense that
machine-dependent code has slightly~decreased.

\begin{table*}[th!b]
\centering
\caption{Logistic Growth Rate Parameters and Their Standard Errors (SEs)$^1$}
\label{tab: logistic}
\begin{threeparttable}
\begin{tabular}{llrrlrrlrrlrr}
\toprule
&& \multicolumn{2}{c}{$\textmd{FILE}_{it}$}
&& \multicolumn{2}{c}{$\textmd{LOC}_{it}$}
&& \multicolumn{2}{c}{$\textmd{BUG}_{it}$}
&& \multicolumn{2}{c}{$\textmd{WARN}_{it}$} \\
\cmidrule{3-4}
\cmidrule{6-7}
\cmidrule{9-10}
\cmidrule{12-13}
&& $\hat{\beta}$ & SE($\hat{\beta}$)
&& $\hat{\beta}$ & SE($\hat{\beta}$)
&& $\hat{\beta}$ & SE($\hat{\beta}$)
&& $\hat{\beta}$ & SE($\hat{\beta}$) \\
\hline
Subsystem & arch & $<$0.0001 & $<$0.0001 && -- & -- && -- & -- && 0.0534 & 0.0015 \\
& drivers & 0.0739 & 0.0028 && 0.0648 & 0.0026 && -- & --  && 0.0676 & 0.0027 \\
& fs & 0.0306 & 0.0033 && 0.0413 & 0.0028 && $<$0.0001 & $<$0.0001 && 0.0553 & 0.0033 \\
& net & 0.0781 & 0.0044 && 0.0641 & 0.0024 && $<$0.0001 & $<$0.0001 && 0.0606 & 0.0020 \\
& sound & 0.0574 & 0.0028 && 0.0594 & 0.0022 && -- & -- && 0.2722 & 0.0336 \\
& others & 0.0649 & 0.0011 && 0.0588 & 0.0016 && 0.0677 & 0.0060 && 0.0580 & 0.0018 \\
\bottomrule
\end{tabular}
\begin{tablenotes}
\begin{scriptsize}
\item{$^1$~Estimates are not shown in case a standard error is larger than a
  parameter estimate.}
\end{scriptsize}
\end{tablenotes}
\end{threeparttable}
\end{table*}

\begin{table*}[th!b]
\centering
\caption{Generalized M-fluctuation Tests ($p$-values)}
\label{tab: m-fluctuation tests}
\begin{tabular}{llcrrrcrrr}
\toprule
&& Series & $\textmd{BUG}_{it}$
& $\overline{\textmd{BUG}}_{it}$
& $\widetilde{\textmd{BUG}}_{it}$ &&
$\textmd{WARN}_{it}$
& $\overline{\textmd{WARN}}_{it}$
& $\widetilde{\textmd{WARN}}_{it}$ \\
&& Model & Poisson & OLS & OLS && Poisson & OLS & OLS \\
\hline
Subsystem & \texttt{arch} && $0.0003$ & $< 0.0001$ & $< 0.0001$
&& $< 0.0001$ & $< 0.0001$ & $< 0.0001$ \\
& \texttt{drivers} && $0.0002$ & $< 0.0001$ & $< 0.0001$
&& $< 0.0001$ & $< 0.0001$ & $< 0.0001$ \\
& \texttt{fs} && $< 0.0001$ & $< 0.0001$ & $< 0.0001$
&& $< 0.0001$ & $< 0.0001$ & $< 0.0001$ \\
& \texttt{net} && $< 0.0001$ & $< 0.0001$ & $< 0.0001$
&& $< 0.0001$ & $< 0.0001$ & $< 0.0001$ \\
& \texttt{sound} && $< 0.0001$ & $< 0.0001$ & $< 0.0001$
&& $< 0.0001$ & $< 0.0001$ & $< 0.0001$ \\
& \texttt{others} && $< 0.0001$ & $< 0.0001$ & $< 0.0001$
&& $< 0.0001$ & $< 0.0001$ & $0.0001$ \\
\bottomrule
\end{tabular}
\end{table*}

\begin{figure*}[p!]
\centering
\includegraphics[width=\linewidth, height=8cm]{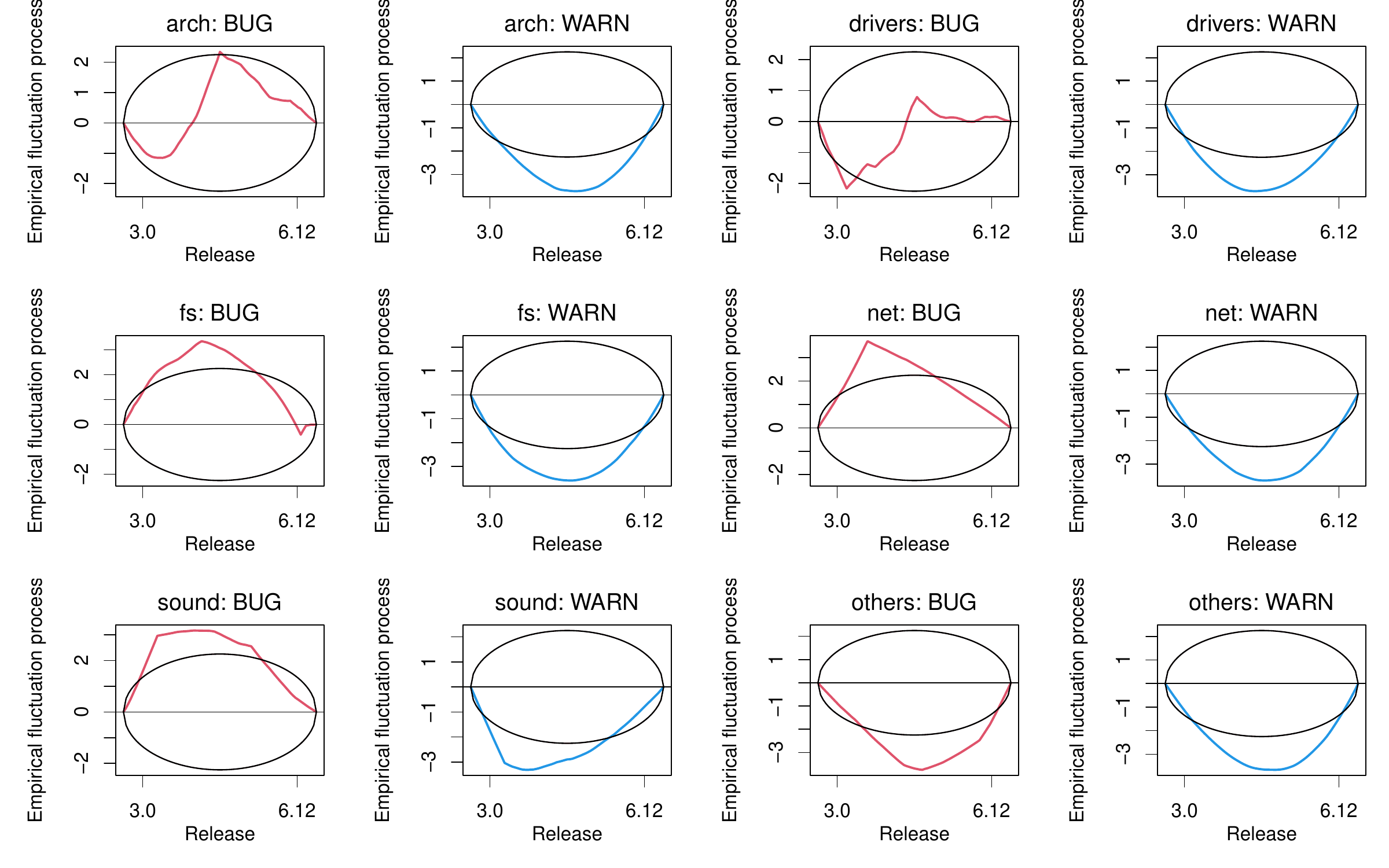}
\caption{OLS-CUSUM Processes for $\textmd{FILE}_{it}$ and $\textmd{BUG}_{it}$
  (the circled boundaries denote a $p < 0.001$ confidence level)}
\label{fig: cusum static}
%\end{figure*}
%
\vspace{20pt}
%
%\begin{figure*}[th!b]
\centering
\includegraphics[width=\linewidth, height=8cm]{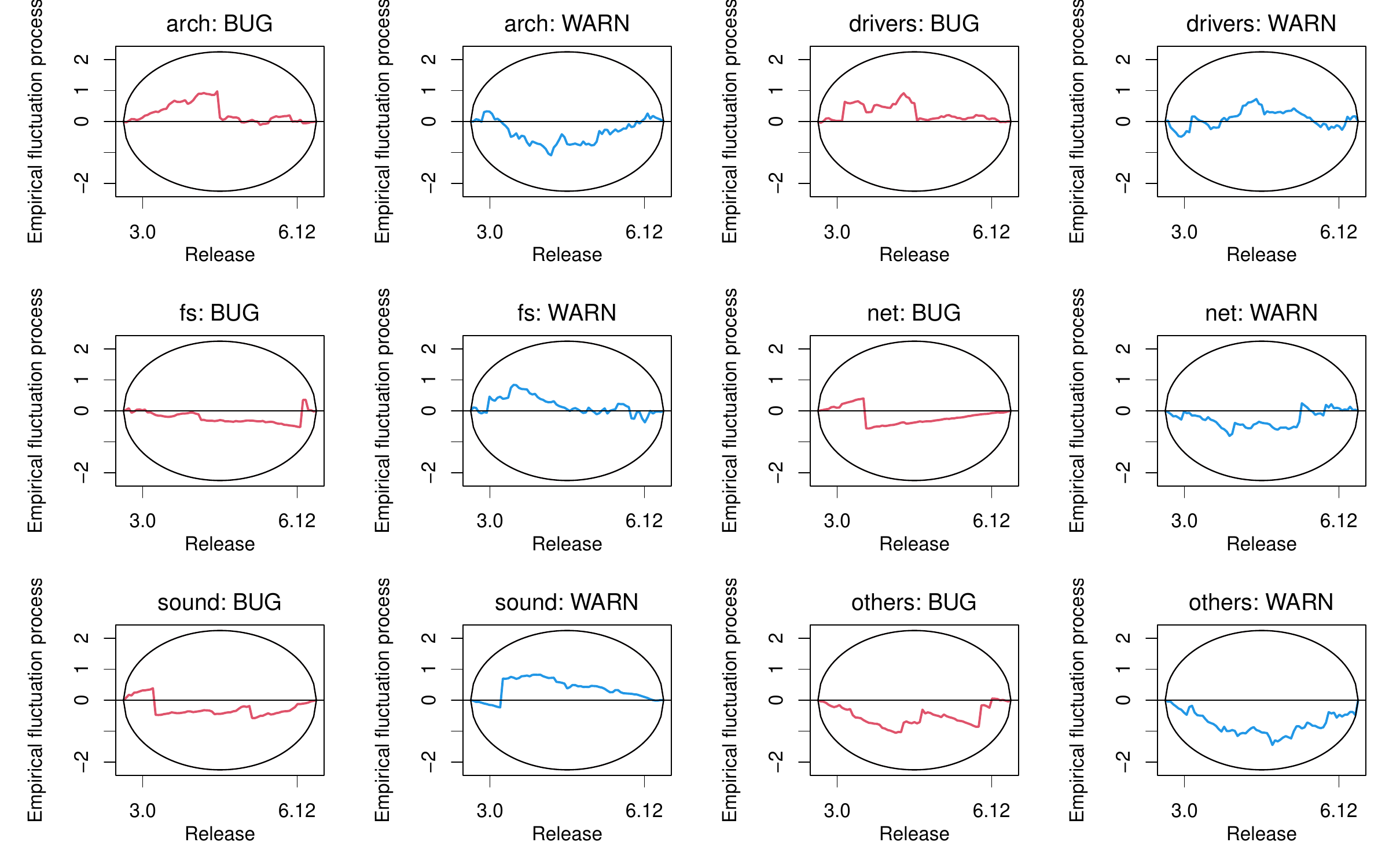}
\caption{OLS-CUSUM Processes for AR(1) models of $\Delta\textmd{FILE}_{it}$ and $\Delta\textmd{BUG}_{it}$ (the boundaries denote a $p < 0.001$ confidence level)}
\label{fig: cusum dynamic}
\end{figure*}

With these observations about growing sizes of most subsystems, the logistic
growth curve estimates can be subsequently considered. The growth rate
parameters estimated are thus shown in Table~\ref{tab: logistic}. As was noted
in the previous Section~\ref{subsec: methods}, some of the estimates are not
reasonable, but regarding those that are, it is important to emphasize that the
coefficients for $\textmd{WARN}_{it}$ are in the majority of subsystems well in
line with those estimated for the two time series $\textmd{FILE}_{it}$ and
$\textmd{LOC}_t$ measuring software size. In fact, the growth rates of soft
assertion use have in some subsystems surpassed the growth rates of software
size. Furthermore, in the \texttt{fs} subsystem and particularly in the
\texttt{net} subsystem the ratio $\textmd{WARN}_{it}~/~\textmd{FILE}_{it}$ has
exceeded a value one. In general, the answer to RQ.1 is positive; in most major
subsystems the growing sizes have been met with equally growing uses of soft
assertions. Having said that, it remains a subject of a debate whether the
growth rates of $\textmd{WARN}_{it}$ should be even higher in order to satisfy
the ever-growing sizes of some subsystems, the implicit design by contract
principle, and the overall goal of providing non-fatal debugging information.

Turning to the CUSUM processes, the formal M-fluctuation tests are summarized in
Table~\ref{tab: m-fluctuation tests}. It can be concluded that all static mean
models reject the null hypotheses of stable processes, no matter of a particular
statistical significance level specified or preferred. When recalling the
negation discussed in Section~\ref{subsec: methods}, the result is a positive
finding with respect to both RQ.1 and RQ.2. In other words, neither the use of
hard assertions nor the use of soft assertions has remained stable over the
years. However, taking a visual look at the underlying CUSUM processes reveals
two exceptions.

As can be seen from Fig.~\ref{fig: cusum static}, the static mean models for the
$\textmd{BUG}_{it}$ time series mostly stay within the circled boundaries in the
\texttt{arch} and \texttt{drivers} subsystems. Only minor boundary crossings are
present for these two time series. In all other subsystems the boundaries are
heavily crossed by the static mean models for both $\textmd{BUG}_{it}$ and
$\textmd{WARN}_{it}$. Although there is no right or wrong answer to a question
whether one should prefer the visual inspection over the formal M-fluctuation
tests, it can be concluded that the answer to RQ.2 is not entirely
uniform. Given the question's phrasing ``some or all'', the conclusion is on the
side of the former word. This conclusion warrants a brief speculation on what
might explain the stable use of hard assertions in the two outlying subsystems
across time. A~reasonable presumption might be that both \texttt{arch} and
\texttt{drivers} operate close to hardware, and thus implementing recovery
routines may be particularly difficult in these subsystems. There may also be
bugs in hardware, and existing observations indicate that many Linux device
drivers tolerate such bugs poorly~\cite{Kadav09}. The \texttt{drivers} subsystem
has also been observed to be the most bug-prone one, including with respect to
regression bugs~\cite{Chou01, Ruohonen24EASE}. To that end, it might also be that
hard assertions are used in the device driver subsystem to annotate code against
verified and fixed bugs.

Finally, the dynamic AR(1) models too deserve a brief exposition. The CU\-SUM
processes for these are visualized in Fig.~\ref{fig: cusum dynamic}. As can be
seen, all fluctuation processes remain neatly within the boundaries. This
observation reflects a fact that the differenced series are stationary; the
means of the release-to-release change rates are constant. The coefficients for
the lagged and differenced series are also modest in magnitude; with an
exception of \texttt{others}, the coefficients are smaller than $0.2$ in
absolute value. All in all, the dynamic AR(1) models signify that no major
refactoring periods involving assertions have occurred. When backtracking to
Fig.~\ref{fig: series}, the only exception is the sudden decline of hard
assertions in the \texttt{net} subsystem late during the kernel's third release
series.

\section{Conclusion}\label{sec: conclusion}

The paper examined the longitudinal evolution of assertions in the Linux kernel
from its third release series to the ongoing sixth release series. Two research
questions were postulated. The answers to these can be summarized as
follows. With respect to RQ.1, the conclusion is that the kernel's continuous
growth, as being driven particularly by the addition of new device drivers, has
been met with a growing use of softer assertion variants that produce warnings
but do not cause an outright kernel panic. Regarding RQ.2, the conclusion is
that at the same time the use of harder, panic-causing assertions has slowly
decreased in some of the kernel's major subsystems. The notable exceptions are
the \texttt{arch} and \texttt{drivers} subsystems together with a leftover
subsystem group labeled as \texttt{others}. In these subsystems the use of the
harder \texttt{BUG()}-style assertions has remained stable or even
increased. Nevertheless, it seems that a slow but visible trend is occurring
in other Linux kernel subsystems toward handling errors in a more graceful way than causing~a~panic.

The handling of errors in the kernel space would also deserve more
research. Indeed, whether, how, and how well an operating system kernel handles
bugs and fatal errors are classical research questions~\cite{Chou01,
  Denning76}. While there is a whole research branch on fault-tolerant operating
systems and their kernels, it would be worthwhile to study also the error
recovery strategies used in existing operating system kernels. The topic could
be studied also empirically by focusing on frameworks and interfaces as well as
programming patterns and idioms used for error handling in the kernel space. A
comparative setup for empirical research could be constructed by again
concentrating on the subsystems of the Linux kernel or enlarging the scope to
other open source software operating system kernels. An interesting research
path would open also by focusing on the potential differences between error
handling with the C and Rust programming languages. Another option would be to
continue with the themes noted in Section~\ref{sec: related work}. For instance,
the existing knowledge seems limited with respect to the alleged benefits of
assertions for documenting code and helping debugging in the kernel
context. Answers to these and related questions might help at reaching the goal
of moving toward software assertion use in the Linux kernel.

\bibliographystyle{splncs03}
%\bibliography{assertlinux}

\begin{thebibliography}{10}
\providecommand{\url}[1]{\texttt{#1}}
\providecommand{\urlprefix}{URL }

\bibitem{Bombieri25}
Bombieri, N., Germiniani, S., Lumpp, F., Pravadelli, G.: {E}dge-{C}loud
  {O}rchestration of {A}ssertion-{B}ased {M}onitors for {R}obotic
  {A}pplications. ACM Transactions on Embedded Computing Systems pp. 1--21
  (2025), published online in March

\bibitem{FreeBSD24}
Bresler, J.M., Horne, M.: {KASSERT} -- {K}ernel {E}xpression {V}erification
  {M}acros (2024), {F}ree{BSD} {M}anual {P}ages, available online in December
  2024: \url{https://man.freebsd.org/cgi/man.cgi?KASSERT(9)}

\bibitem{Brown75}
Brown, R.L., Durbin, J., Evans, J.M.: {T}echniques for {T}esting the
  {C}onstancy of {R}egression {R}elationships over {T}ime. Journal of the Royal
  Statistical Society. Series B (Methodological)  37(2),  149--192 (1975)

\bibitem{Chen12}
Chen, B., Hong, Y.: {T}esting for {S}mooth {S}tructural {C}hanges in {T}ime
  {S}eries {M}odels via {N}onparametric {R}egression. Econometrica  80(3),
  1157--1183 (2012)

\bibitem{Chen24}
Chen, Z., Jia, C., Chen, L.: {E}valuating {T}est {Q}uality of {P}ython
  {L}ibraries for {IoT} {A}pplications at the {N}etwork {E}dge. Wireless
  Networks  30,  6603--6618 (2024)

\bibitem{Chou01}
Chou, A., Yang, J., Chelf, B., Hallem, S., Engler, D.: {A}n {E}mpirical {S}tudy
  of {O}perating {S}ystems {E}rrors. In: Proceedings of the Eighteenth ACM
  Symposium on Operating Systems Principles (SOSP 2001). pp. 73--88. ACM, Banff
  (2001)

\bibitem{Clarke06}
Clarke, L.A., Rosenblum, D.S.: {A} {H}istorical {P}erspective on {R}untime
  {A}ssertion {C}hecking in {S}oftware {D}evelopment. ACM SIGSOFT Software
  Engineering Notes  31(3),  25--37 (2006)

\bibitem{LWN24a}
Corbet, J.: {W}arning about {WARN\_ON()} (2024), {L}inux {W}eekly {N}ews
  ({LWN}). Available online in December 2024:
  \url{https://lwn.net/Articles/969923/}

\bibitem{Denning76}
Denning, P.J.: {F}ault {T}olerant {O}perating {S}ystems. ACM Computing Surveys
  8(4),  359--389 (1976)

\bibitem{Gu03}
Gu, W., Kalbarczyk, Z., Iyer, R.K., Yang, Z.: {C}haracterization of {L}inux
  {K}ernel {B}ehavior under {E}rrors. In: Proceedings of the 43rd Annual
  IEEE/IFIP International Conference on Dependable Systems and Networks (DSN
  2003). pp. 459--469. IEEE (2003)

\bibitem{Hoare03}
Hoare, C.A.R.: {A}ssertions: {A} {P}ersonal {P}erspective. IEEE Annals of the
  History of Computing  25(2),  14--25 (2003)

\bibitem{Israeli10}
Israeli, A., Feitelson, D.G.: {T}he {L}inux {K}ernel as a {C}ase {S}tudy in
  {S}oftware {E}volution. Journal of Systems and Software  83(3),  485--501
  (2010)

\bibitem{JiangJiang24}
Jiang, M., Jiang, J., Wu, T., Ma, Z., Luo, X., Zhou, Y.: {U}nderstanding
  {V}ulnerability {I}nducing {C}ommits of the {L}inux {K}ernel. ACM
  Transactions on Software Engineering and Methodology  33,  170:1 -- 170:28
  (2024)

\bibitem{Kadav09}
Kadav, A., Renzelmann, M.J., Swift, M.M.: {T}olerating {H}ardware {D}evice
  {F}ailures in {S}oftware. In: Proceedings of the ACM SIGOPS 22nd Symposium on
  Operating Systems Principles (SOSP 2009). pp. 59--72. ACM, Big Sky Montana
  (2009)

\bibitem{Kudrjavets06}
Kudrjavets, G., Nagappan, N., Ball, T.: {A}ssessing the {R}elationship between
  {S}oftware {A}ssertions and {F}aults: {A}n {E}mpirical {I}nvestigation. In:
  Proceedings of the 17th International Symposium on Software Reliability
  Engineering (ISSRE 2006). pp. 204--212. Raleigh (2006)

\bibitem{Lehman97}
Lehman, M.M., Ramil, J.F., Wernick, P.D., Perry, D.E., Turski, W.M.: {M}etrics
  and {L}aws of {S}oftware {E}volution---{T}he {N}ineties {V}iew. In:
  Proceedings Fourth International Software Metrics Symposium. pp. 20--32.
  Albuquerque (1997)

\bibitem{Meyer92}
Meyer, B.: {A}pplying ``{D}esign by {C}ontract''. Computer  25(10),  40--51
  (1992)

\bibitem{Muller02}
M\"uller, M.M., Typke, R., Hagner, O.: {T}wo {C}ontrolled {E}xperiments
  {C}oncerning the {U}sefulness of {A}ssertions as a {M}eans for {P}rogramming.
  In: Proceedings of the International Conference on Software Maintenance (ICSM
  2002). pp. 84--92. IEEE, Montreal (2002)

\bibitem{Olufowobi19}
Olufowobi, H., Ezeobi, U., Muhati, E., Robinson, G., Young, C., Zambreno, J.,
  Bloom, G.: {A}nomaly {D}etection {A}pproach {U}sing {A}daptive {C}umulative
  {S}um {A}lgorithm for {C}ontroller {A}rea {N}etwork. In: Proceedings of the
  ACM Workshop on Automotive Cybersecurity (AutoSec 2019). pp. 25--30. ACM,
  Richardson

\bibitem{Patel17}
Patel, K., Hierons, R.M.: {A} {M}apping {S}tudy on {T}esting {N}on-{T}estable
  {S}ystems. Software Quality Journal  26,  1373--1413 (2018)

\bibitem{Peruma24}
Peruma, A., Takebayashi, T., Huang, R., Averion, J.C., Hodapp, V., Newman,
  C.D., Mkaouer, M.W.: {O}n the {R}ationale and {U}se of {A}ssertion {M}essages
  in {T}est {C}ode: {I}nsights from {S}oftware {P}ractitioners. In: Proceedings
  of the IEEE International Conference on Software Maintenance and Evolution
  (ICSME 2024). pp. 538--549. IEEE, Flagstaff (2024)

\bibitem{Rosenblum92}
Rosenblum, D.S.: {T}owards a {M}ethod of {P}rogramming {W}ith {A}ssertions. In:
  Proceedings of the 14th International Conference on Software Engineering
  (ICSE 1992). pp. 92--10. ACM, Melbourne (1992)

\bibitem{Rosenblum95}
Rosenblum, D.S.: {A} {P}ractical {G}uide to {P}rogramming {W}ith {A}ssertions.
  IEEE Transactions on Software Engineering  21(1),  19--31 (1995)

\bibitem{Ruohonen24rep}
Ruohonen, J.: {A} {R}eplication {P}ackage for a {P}aper {E}ntitled ``{A} {T}ime
  {S}eries {A}nalysis of {A}ssertions in the {L}inux {K}ernel'' (2024),
  {H}arvard {D}ataverse. {A}vailable online in December 2024:
  \url{https://doi.org/10.7910/DVN/TFBVPJ}

\bibitem{Ruohonen24EASE}
Ruohonen, J., Alami, A.: {F}ast {F}ixes and {F}aulty {D}rivers: {A}n
  {E}mpirical {A}nalysis of {R}egression {B}ug {F}ixing {T}imes in the {L}inux
  {K}ernel (2024), {A}rchived manuscript, available online:
  \url{https://arxiv.org/abs/2411.02091}

\bibitem{Ruohonen15COSE}
Ruohonen, J., Hyrynsalmi, S., Lepp\"anen, V.: {T}he {S}igmoidal {G}rowth of
  {O}perating {S}ystem {S}ecurity {V}ulnerabilities: {A}n {E}mpirical
  {R}evisit. Computers \& Security  55,  1--20 (2015)

\bibitem{Ruohonen15JSEP}
Ruohonen, J., Hyrynsalmi, S., Lepp\"anen, V.: {T}ime {S}eries {T}rends in
  {S}oftware {E}volution. Journal of Software: Evolution and Process  27(2),
  990--1015 (2015)

\bibitem{Ruohonen19RSDA}
Ruohonen, J., Rindell, K.: {E}mpirical {N}otes on the {I}nteraction {B}etween
  {C}ontinuous {K}ernel {F}uzzing and {D}evelopment. In: Proceedings of the
  IEEE International Symposium on Software Reliability Engineering Workshops
  (ISSREW 2019). pp. 276--281. IEEE, Berlin (2019)

\bibitem{Sprouffske16}
Sprouffske, K., Wagner, A.: {G}rowthcurver: {A}n {R} {P}ackage for {O}btaining
  {I}nterpretable {M}etrics {F}rom {M}icrobial {G}owth {C}urves. BMC
  Bioinformatics  19,  1--4 (2016)

\bibitem{Taromirad24}
Taromirad, M., Runeson, P.: {A} {L}iterature {S}urvey of {A}ssertions in
  {S}oftware {T}esting. In: Proceedings of the 8th International Conference on
  Engineering of Computer-Based Systems (ECBS 2023). pp. 75--96. Springer,
  V\"aster\r{a}s (2024)

\bibitem{Teto17}
Teto, J.K., Bearden, R., Lo, D.C.T.: {T}he {I}mpact of {D}efensive
  {P}rogramming on {I/O} {C}ybersecurity {A}ttacks. In: Proceedings of the 2017
  ACM Southeast Conference (ACMSE 2017). pp. 102--111. ACM, Kennesaw (2017)

\bibitem{Kernel24b}
{The Kernel Development Community}: {H}ow the {D}evelopment {P}rocess {W}orks
  (2024), the Linux Kernel Documentation. Available online in December 2024:
  \url{https://docs.kernel.org/process/2.Process.html}

\bibitem{Kernel24a}
{The Kernel Development Community}: {L}inux {K}ernel {C}oding {S}tyle (2024),
  the Linux Kernel Documentation. Available online in December 2024:
  \url{https://docs.kernel.org/process/coding-style.html}

\bibitem{Kernel24c}
{The Linux Kernel Organization, Inc.}: {T}he {L}inux {K}ernel {A}rchives
  (2024), available online in December 2024:
  \url{https://www.kernel.org/pub/linux/kernel/}

\bibitem{Winkler24}
Winkler, D., Urbanke, P., Ramler, R.: {I}nvestigating the {R}eadability of
  {T}est {C}ode. Empirical Software Engineering  29,  1--48 (2024)

\bibitem{Witharana25}
Witharana, H., Weerasena, H., Mishra, P.: {S}ecurity {A}ssertions for {T}rusted
  {E}xecution {E}nvironments. In: Proceedings of the Design, Automation \& Test
  in Europe Conference (DATE 2025). pp. 1--6. IEEE, Lyon (2025)

\bibitem{Yoshimura12}
Yoshimura, T., Yamada, H., Kono, K.: {I}s {L}inux {K}ernel {O}ops {U}seful or
  {N}ot? In: Proceedings of the Eighth Workshop on Hot Topics in System
  Dependability (HotDep 2012). pp. 1--6. USENIX, Hollywood (2012)

\bibitem{Zeileis06}
Zeileis, A.: {I}mplementing a {C}lass of {S}tructural {C}hange {T}ests: {A}n
  {E}conometric {C}omputing {A}pproach. Computational Statistics \& Data
  Analysis  50(11),  2987--3008 (2006)

\bibitem{Zeileis02}
Zeileis, A., Leisch, F., Hornik, K., Kleiber, C.: strucchange: {A}n {R}
  {P}ackage for {T}esting for {S}tructural {C}hange in {L}inear {R}egression
  {M}odels. Journal of Statistical Software  7(2),  1--38 (2002)

\bibitem{Zhang15}
Zhang, Y., Mesbah, A.: {A}ssertions {A}re {S}trongly {C}orrelated {W}ith {T}est
  {S}uite {E}ffectiveness. In: Proceedings of the 2015 10th Joint Meeting on
  Foundations of Software Engineering (ESEC/FSE 2015). pp. 214--224. ACM,
  Bergamo (2015)

\end{thebibliography}

\end{document}